\documentclass[reprint,pra,longbibliography,superscriptaddress,nofootinbib]{revtex4-2}

\usepackage{lipsum}
\usepackage{fancyhdr}
\usepackage{graphicx}
\usepackage[caption=false]{subfig}
\usepackage{braket}
\usepackage{booktabs}
\usepackage{multirow}
\usepackage{bm}
\usepackage{mathtools}
\usepackage{amsmath, esint}
\usepackage{amssymb}
\usepackage{hyperref}
\usepackage{esvect}
\usepackage{float}
\usepackage{placeins}
\usepackage{verbatim}
\usepackage{epstopdf}
\usepackage[normalem]{ulem}
\usepackage{mwe}
\usepackage{array}
\usepackage{soul}

\hypersetup{
    colorlinks=true,
    linkcolor=blue,
    filecolor=magenta,      
    urlcolor=cyan,
    citecolor={blue},
    }

\newcommand{\bl}[1]{{\color{blue}#1}}

\allowdisplaybreaks

\makeatletter

\usepackage{comment}
\let\wfs@comment@comment\comment
\let\comment\@undefined

\usepackage[final]{changes}
\@namedef{Changes@AuthorColor}{red}
\colorlet{Changes@Color}{red}

\let\wfs@changes@comment\comment
\let\comment\@undefined

\newcommand\comment{%
    \ifthenelse{\equal{\@currenvir}{comment}}
    {\wfs@comment@comment}
    {\wfs@changes@comment}%
}

\makeatother

\begin{document}

\title{Reducing Disorder-Induced Backscattering 
in Photonic Crystal Waveguides through Inverse Design}

\author{Dominic Thompson}
\email{19djt@queensu.ca}
\affiliation{Centre for Nanophotonics, Department of Physics, Engineering Physics and Astronomy, Queen's University, Kingston, Ontario, Canada, K7L 3N6}

\author{Antonia Neill}
\affiliation{Centre for Nanophotonics, Department of Physics, Engineering Physics and Astronomy, Queen's University, Kingston, Ontario, Canada, K7L 3N6}

\author{Nir Rotenberg}
\affiliation{Centre for Nanophotonics, Department of Physics, Engineering Physics and Astronomy, Queen's University, Kingston, Ontario, Canada, K7L 3N6}

\author{Stephen Hughes}
\affiliation{Centre for Nanophotonics, Department of Physics, Engineering Physics and Astronomy, Queen's University, Kingston, Ontario, Canada, K7L 3N6}

\begin{abstract}
Photonic crystal waveguides (PCWs) allow for the engineering of photonic waveguide modes and band structures to control the flow of light and enhance light-matter interactions within the waveguide. They have shown potential for enhancing optical nonlinearities, single photon emissions from quantum dots, as well as for use in optical buffers due to their ability to confine fields on-chip and produce slow-light modes. While these features are promising for applications in nanophotonics, PCWs are prone to high scattering losses due to disorder-induced backscattering, which has remained a major challenge across various waveguide designs for decades. By combining a fast mode solving approach with physics-based scattering formulas and inverse design, we show how backscattering losses can be significantly reduced, even when working at the same group index. We demonstrate substantial improvements for both W1-like waveguide modes and topological waveguide modes. Our general methodology is fully three-dimensional and can be used to design new PCWs optimized for a variety of performance metrics. 
\end{abstract}

\maketitle

\section{Introduction }

Photonic crystal waveguides (PCWs) have 
shown great promise for enhancing various
optical properties since the late 1980s \cite{Yablonovitch1987,John1987,PHCBook,Joannopoulos1997,Krauss2008,Notomi2001}. 
In particular, PCWs can be dispersion-engineered to create 
{\it slow-light} mode regions, where the reduced velocity of light can, for example, enhance optical nonlinearities \cite{Soljacic:02} and interactions with semiconductor quantum dots \cite{Hughes2004,MangaRao2007,RevModPhys.87.347}, as well as be used for all-optical switches \cite{Beggs:08}. However, this increased light-matter interaction strength also enhances disorder-induced backscattering \cite{Hughes2005}, which has a major impact on practical applications. 

Disorder-induced backscattering losses, to a first approximation, scale quadratically with the mode group index (the {\it slowdown factor}, $n_g$), though many beneficial optical effects only scale linearly \cite{OFaolain2007}. This greatly limits the usefulness of PCWs for slow-light applications. However, the general scaling of backscatter is {\it not} simply proportional to $n_g^2$, as it also depends on the strength of the electric field at the PCW hole edges 
\cite{Hughes2005}. The scaling can also change for longer samples due to multiple scattering events \cite{pattersonDisorderinducedIncoherentScattering2009}, but we will not explore that here.
The problem of disorder-induced backscattering in PCWs has been an active area of research for over 20 years \cite{Krauss2008,Hughes2005,pattersonDisorderinducedIncoherentScattering2009}. A few designs have been found to only modestly reduce backscattering, with the best case being around a factor of two for the same quality waveguides and fabrication \cite{Notomi:07,Mann:13,Wang:12}.

\begin{figure*}[t]
    \centering
    \includegraphics[width=.9\textwidth]{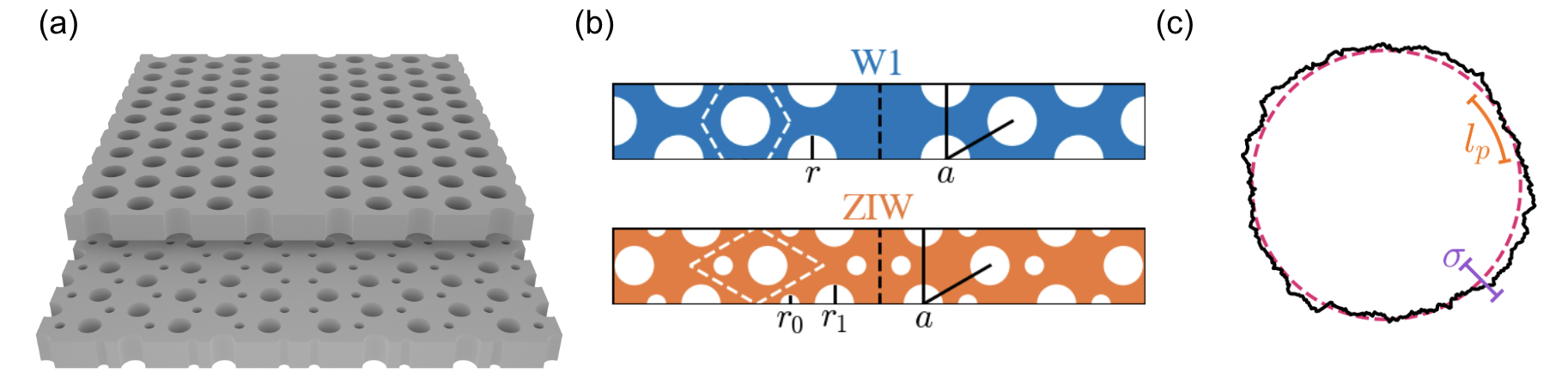}
    \caption{(a) A 3D view and (b) a unit cell of the W1 and ZIW waveguides on the top and bottom, respectively. In this paper, $a=266$ nm, $r=79.8$ nm, $r_0=27.9$ nm, $r_1=62.5$ nm, and $h=170$ nm, where $h$ is the thickness of the slab. The slab material has a permittivity of $\epsilon_2=3.4638$, and the surrounding material is air with a permittivity of $\epsilon_1=1$. The dashed vertical lines indicate the direction of propagation. (c) A representation of the type of fabrication noise assumed in this paper. The statistical parameters are the standard deviation, $\sigma$, and the correlation length, $l_p$. In this paper, we use $l_p=40$ nm and $\sigma=3$ nm.}
    \label{fig:diagram}
\end{figure*}

Recently, topological PCWs have been proposed to potentially help with the backscattering problem, inspired by topological electronics, where losses have been reduced through the breaking of time reversal symmetry and the quantum Hall effect \cite{Lu2016}. In electromagnetism, time reversal symmetry can be broken using magneto-optical materials in GHz frequencies \cite{Wang2009}, but these materials are inherently lossy and challenging to manufacture in the optical regime. Alternate designs demonstrated topologically non-trivial properties at optical frequencies by relying on the crystal symmetries that limit coupling between forward and backward propagating modes \cite{Barik2016,Anderson:17}, but these modes are above the light line, and thus have large radiative (intrinsic) losses \cite{Sauer2020}. 
The current state-of-the-art topological PCWs improve upon these designs 
by using the {\it valley Hall} effect \cite{Ma2016,Shalaev2018,He2019},
and 
bringing the modes below the light line.  This was thought to potentially reduce disorder-induced backscattering, and these designs have shown promise in increasing transmission around bends in the PCW \cite{jalali_mehrabad_chiral_2020};
however, the general arguments for reduced losses are not well understood and are not generally applicable, and recent experiments and simulations have shown that they are still heavily affected by disorder-induced backscattering \cite{Rosiek2023,Hauff2022}.

As highlighted above,
disorder-induced backscattering is not solely dependent on the group index of light or the amount of disorder, but also depends on the distribution of the electric field Bloch mode, and some PCWs designed with smaller loss have 
benefited from this fact \cite{Notomi:07,Mann:13,Wang:12,Petrov:09}. Specifically, backscattering is increased when a larger field is located along the edge of the hole, exactly where the fabrication disorder occurs \cite{Hughes2005}. Since a large group index is key to other desirable properties \cite{Krauss2008}, a general design goal is to reduce backscattering without having a large effect on the group index. In this work, we address this challenge by using gradient-based inverse design. Importantly, we do this 
using an efficient and fully 3D, semi-analytical mode-solving approach, the guided mode expansion (GME) method; this makes the problem tractable as a full 3D approach, while a brute force solution of Maxwell's equations would be unfeasible due to the computational demands.

Inverse design has become increasingly prevalent in photonics \cite{Nussbaum2021,PhysRevA.106.033514,Christiansen:21,Molesky2018,Deng2024,PhysRevResearch.6.L022065}, and while there are a variety of inverse design frameworks including heuristic and machine learning methods, here we  use gradient based inverse design, since it is very efficient when gradients for the function can be found~\cite{Minkov2020,Deng2024}. Specifically, we use the open-source Python library {\it Legume}, which allows GME to efficiently calculate full 3D mode solutions for photonic crystal slabs, and compute gradients of the modes through automatic differentiation~\cite{Zanotti2024}. The GME is an approximate method when truncated in basis size, but it is known to be highly accurate for high-index contrast PCWs \cite{PhysRevB.73.235114,Crane2017,Vasco2018,Vasco2019}, and is much more efficient than using full Maxwell solvers that employ traditional approaches such as finite-elements (e.g., COMSOL) and finite-difference time-domain (FDTD) methods due to its far superior speed \cite{PhysRevB.73.235114}---{\it typically seconds versus hours on a standard desktop computer}.

To demonstrate the general power of our approach, {\it we show how to 
significantly improve (i.e., reduce) the disorder-induced backscattering loss in both the W1-like and ZIW-like (zigzag interface) waveguides} as shown schematically in Fig.~\ref{fig:diagram}(a-b). The W1 is a well-studied design with a single row of missing holes, while the ZIW waveguide is one of the most promising topological PCW based on the valley Hall effect \cite{Hauff2022}.

The rest of our paper is organized as follows: 
in Sec.~\ref{sec: methods}
we first explain the methods used in more detail, starting with GME in Sec.~\ref{sec: GME}, then discuss the backscattering formula in Sec.~\ref{sec:back}, and finally describe the inverse design method in Sec.~\ref{sec:inverse}. We subsequently show two designs for PCWs, where each case
has their loss improved upon, compared to their original design, 
by around an order-of-magnitude (upon removing the trivial group index dependence), and then present two W1-like designs targeting low loss at specific group indices in Sec.~\ref{sec:results}. Finally, we summarize our findings and briefly discuss some perspectives and future work in Sec.~\ref{sec:coclu}.

\section{Theory and Methods}
\label{sec: methods}

In this section, we first briefly discuss the GME method, and explain how it can be used to perform efficient inverse design of disorder-induced backscattering of PCWs. We start by summarizing the GME method, its required assumptions, and the resulting computational speed-up compared to conventional methods that use a full numerical solution of Maxwell's equations. We then introduce the key analytical formula for describing backscattering that we will use throughout our optimizations, which we emphasize has been used to explain various experiments. Finally, we describe the optimization method we employ for the inverse design, with a view to reducing backscattering, while retaining a well-defined single mode within the photonic bandgap and below the light line. This ensures that the nominal mode 
of interest remains lossless in the absence of disorder.

\subsection{Guided Mode Expansion} \label{sec: GME}
The GME method solves for the photonic bands and mode structure as an eigenvalue problem for a given wave vector within of a photonic crystal dielectric slab, and also returns the field Bloch modes. 

Assuming that the dielectric material is lossless and frequency-independent in the frequency range of interest, as is typical for semiconductors used at optical frequencies, the magnetic field modes are obtained from the following eigenvalue equation,
\begin{equation} \label{eq:evProb}
    \bm\nabla\times\left(\frac{1}{\epsilon(\bm r)}\bm\nabla\times\bm H_{\bm k}(\bm r)\right)=\left(\frac{\omega}{c}\right)^2\bm H_{\bm k}(\bm r),
\end{equation}
subject to Bloch's boundary conditions, where $\bm k$ indicates the wave vector
(within the Brillouin zone) of the mode, and $\epsilon(\bm r)$ is the position-dependent dielectric constant. 
The structure for the slabs can be seen in 
Fig.~\ref{fig:diagram}(a).

To solve this GME eigenvalue problem, one can expand the magnetic field in a basis of the guided modes of the equivalent homogeneous slab, i.e., a slab with a spatially averaged dielectric constant.
This has a significant advantage, as 
 the field distributions of these modes can be solved analytically. We work in the frequency domain, with the modes indexed by their wave vector $\bm k$, so we 
can define the analytical form for the modes of the {\it inhomogeneous slab} as $\bm H_{\bm G+\bm k,\beta}^{\text{guided}}(\bm r)$, where $\beta$ is the mode index at the specified $\bm G+\bm k$ point, and the $\bm G$s are the reciprocal lattice vectors corresponding to the homogeneous slab unit cell defined during computation. 

For simplicity, one can index these modes as $(\bm G,\beta)\rightarrow\mu$. With this definition, the true photonic crystal slab modes can be written as \cite{PhysRevB.73.235114}
\begin{equation} \label{eq:expansion}
    \bm H_{\bm k}(\bm r)=\sum_\mu c_\mu\bm H_{\bm k,\mu}^\text{guided}(\bm r).
\end{equation}
For this expansion to be exact, the sum needs to be infinite, and the key approximation of the GME method is that it truncates this sum, thereby producing an incomplete basis. For high index semiconductors, however, only a few guided modes are needed in the expansion, typically two to three at most. In addition, for PCW problems, one has to ensure that the unit cell is large enough to avoid supercell effects in the transverse direction, but this
is straightforward to verify for modes that are inside the photonic bands.

Substituting Eq.~\eqref{eq:expansion} into Eq.~\eqref{eq:evProb}, yields 
\begin{equation}
    \sum_v\mathcal H_{\mu v}c_v=\left(\frac{\omega}{c}\right)^2c_\mu ,
\end{equation}
with 
\begin{equation}
    \mathcal{H}_{\mu v}=\int\frac{1}{\epsilon(\bm r)}\left[\bm \nabla\times\bm H_\mu^{*}(r)\right]\left[\bm \nabla\times\bm H_v(r)\right]d\bm r,
\end{equation}
where the $\bm k$  and ``guided'' labels have been dropped for clarity. Using this formulation, the modes and associated frequencies can be found by numerically computing the eigenvalues and eigenvectors of $\mathcal H$. Subsequently, the electric field modes can be found using
\begin{equation}
    \bm E_{\bm k}(\bm r)=\frac{i}{\omega\epsilon_0\epsilon(\bm r)}\bm \nabla\times\bm H_{\bm k}(\bm r).
\end{equation}

Since the PCW is periodic,
we define the field modes in terms of their Bloch modes,
\begin{align}
    \bm E_{\bm k}(\bm r)={\bf e}_{\bm k}(\bm r)e^{i\bm k\cdot \bm r},\\\bm H_{\bm k}(\bm r)={\bf u}_{\bm k}(\bm r)e^{i\bm k\cdot \bm r},
\end{align}
and we normalize these modes such that $\int_{\text{cell}}\epsilon(\bm r)|{\bf e}_{\bm k}(\bm r)|^2d\bm r=1$ for all $\bm k$ values. 

For the work done here, good numerical convergence was found when including all $|\bm G|<4\  \ (a/2\pi)$ reciprocal lattice vectors ($N_{\bm G}=855$ in total), three guided modes,
and a computational unit cell containing ten crystal unit cells on each side of the waveguide defect region. Using the Python package \textit{Legume} as the implementation of GME \cite{Minkov2020}, each ${\bm k}$ point requires approximately 10 seconds on a standard desktop with an eight-core processor, orders of magnitude faster than solvers such as finite-element (COMSOL) or FDTD methods. Thus, 
{\it the entire band structure can be computed in minutes, using this fully 3D approach}.

\subsection{Basic Theory and Model of Disorder-Induced Backscattering} \label{sec:back}

In a realistic PCW, the edges of the holes will be imperfect and deviate from the perfectly
periodic lattice because of structural disorder. Since the disorder roughness is typically very small compared to the size of the structural features of the PCW (e.g., a few percent), it has been shown that the effects of this roughness can be studied using a perturbation theory approach. This method can then be combined with Green's function theory to obtain an {\it analytical form} for the disorder-induced backscattering, as shown below \cite{Hughes2005,Mann2015,Patterson2010_2}: 
\begin{align} \label{eq:backSmall}
    \langle\alpha_{\text{back}}\rangle_k=&\bigg(\frac{a\omega n_g}{2}\bigg)^2\int\int_{\text{cell}} d\bm{r}d\bm{r}'\langle 
\Delta\epsilon(\bm{r})\Delta\epsilon(\bm{r}')\rangle\\&\times[{\bf e}_{k}(\bm{r})\cdot{\bf e}_{k}(\bm{r})][{\bf e}^*_{k}(\bm{r}')\cdot{\bf e}^*_{k}(\bm{r}')]e^{i2k(x-x')},\notag
\end{align}
where $a$ is the lattice constant [see Fig.~\ref{fig:diagram}~(b)], $\omega$ is the mode's frequency, $k$ is the magnitude of $\bm k$ in the direction of the waveguide, and $n_g=c\frac{dk}{d\omega}$ is the group index. This formula gives the 
disorder-induced loss associated with a nominally lossless waveguide mode below the light line. 

The spatial integral in the loss formula is performed over the unit cell shown in Fig.~\ref{fig:diagram}(b), giving the loss values per unit cell, which can then easily be scaled to a particular length
(e.g. per mm or cm).
Specifically, 
this is the intensity (or power) loss following the standard Beer-Lambert law for disorder-induced backscattering under a thin-sample approximation \cite{Mann:13,Patterson2010}. For longer samples, there will be multiple scattering events, resulting in Anderson localization \cite{Crane2017,Sapienza2010,Vasco2017,pattersonDisorderinducedIncoherentScattering2009}, though coupled mode equations can easily be solved to also obtain expectation values for longer samples \cite{pattersonDisorderinducedIncoherentScattering2009}.

The key term describing the structural disorder is $\langle\Delta\epsilon(\bm{r})\Delta\epsilon(\bm{r}')\rangle$, which defines the ensemble averaged disorder of the dielectric. We assume that this is well described by a statistical function with intra-hole correlation and fluctuations of the hole side walls, a model that has been used with success to explain various experiments \cite{Crane2017,PhysRevLett.102.253903,Mann:13, pattersonDisorderinducedIncoherentScattering2009}. This function is dependent on $\Delta R$, the statistical disorder in the radius of each air hole, so that
\begin{align}\label{eq:noise}
    \Delta\epsilon(\bm r)=&(\epsilon_2-\epsilon_1)\Theta\left(\frac{h}{2}-z\right)\sum_j\Delta R(\tilde \phi)\notag\\&\times \delta\left(R_j-|\bm \rho -\bm \rho_j|\right),
\end{align}
where $j$
is the hole index, $\epsilon_1$ and $\epsilon_2$ are the permittivity values of the slab and surrounding material,
and $\Theta(h/2-z)$ is the heavy-step function that brings all values outside the slab to zero; in addition, $\tilde \phi$ is the polar angle around the hole, $R_j$ is the radius of the hole, $\bm \rho$ is position vector projected into the $x$-$y$ plane, and $\bm \rho_j$ is the $x$-$y$ location of the hole. 

Therefore, the change in the permittivity is localized to the edge of each hole and is proportional to the change in radius caused by fabrication imperfections. We choose the correlation function of the change in radius to be
\begin{align}
    \langle\Delta R(\tilde\phi)\Delta R(\tilde\phi')\rangle = \sigma^2\exp\left(\frac{-R_j|\tilde\phi-\tilde\phi'|}{l_p}\right)\delta_{j,j'} ,   
\end{align}
where $\sigma$ and $l_p$ are the standard deviation and correlation length of the fluctuation noise on the hole edge, respectively. The effects of these parameters are shown in 
Fig.~\ref{fig:diagram}(c). This correlation function has been shown to agree with fabricated structures, with the statistical parameters depending on the fabrication quality \cite{Hauff2022,pattersonDisorderinducedIncoherentScattering2009,Skorobogatiy:05}.


Finally, the full 
backscattering formula,
for use in fast numerical calculations, can then be written as 
\begin{widetext}
\begin{align}\label{eq:backFull}
    \langle\alpha_{\text{back}}\rangle_k
    &=\sum_{j}\bigg(\frac{a\omega_kn_g\sigma}{2}\bigg)^2(\epsilon_2-\epsilon_1)^2\int\int_{\text{cell}}d\bm{r}d\bm{r}'
    \Theta\bigg(\frac{h}{2}-|z|\bigg) \Theta\bigg(\frac{h}{2}-|z'|\bigg)\nonumber \\
    & \times \delta(R_j-|\bm{\rho}-\bm{\rho}_j|) [{\bf e}^*_{k}(\bm{r})\cdot{\bf p}^*_{k}(\bm{r})][{\bf e}_{k}(\bm{r}')\cdot{\bf p}_{k}(\bm{r}')]
    \exp\bigg(\frac{-R_j|\tilde\phi-\tilde\phi'|}{l_p}+i2k(x-x')\bigg),
\end{align}
\end{widetext}
where ${\bf p}_k(\bm r)$ denotes the polarization density,
defined as
\begin{equation}
    {\bf p}_k(\bm r)=\left[{\bf e}_{k,||}(\bm r)+\epsilon(\bm r)\frac{{\bf d}_{k,\perp}(\bm r)}{\epsilon_1\epsilon_2}\right],
\end{equation}
which accounts for local field corrections; note 
that ${\bf e}_{k,||}(\bm r)$ is the component of the electric field parallel to the hole edge, while $d_{k,\perp}(\bm r)$ is the perpendicular component of the displacement field. 
It is important to highlight that due to local-field corrections, the effect of disorder is not symmetrical, and, for example, can also cause disorder-induced frequency shifts of the mode~\cite{PhysRevA.92.023849}.

Thus, from
Eq.~\eqref{eq:backFull}, we see that the backscattering depends primarily on two terms: the group index and the electric field mode around the hole edges. The group index, as previously mentioned, is $n_g=c/{v_g}$, where $v_g$ is the group velocity. The group index is proportional to the inverse of the band curvature and characterizes the speed at which light travels through the waveguide. 
This factor simply provides a scaling to the loss; therefore, 
what matters most is the loss per slowdown factor \bl{squared}. 
For the main design problem, as is made clear from the analytical formula in Eq.~\eqref{eq:backFull}, we are primarily interested in reducing the field around the hole edge, since when a large field interacts with perturbations in the hole edges, it results in backscattering \cite{Mann:13}. 


We highlight that the above loss derivation is 
relevant for regions of the band structure where the frequency is single-valued within the band structure (i.e., single mode). If the frequency is double (or multi) mode, then the multiple modes will couple to one another, leading to additional losses  \cite{Kuramochi2005,Parini2008}. While these additional losses could also be computed, they are not the focus of this present work, and we will also require that the designs of interest remain single mode in a frequency region of interest.

\begin{figure*}[t]
    \centering
    \includegraphics[width=\textwidth]{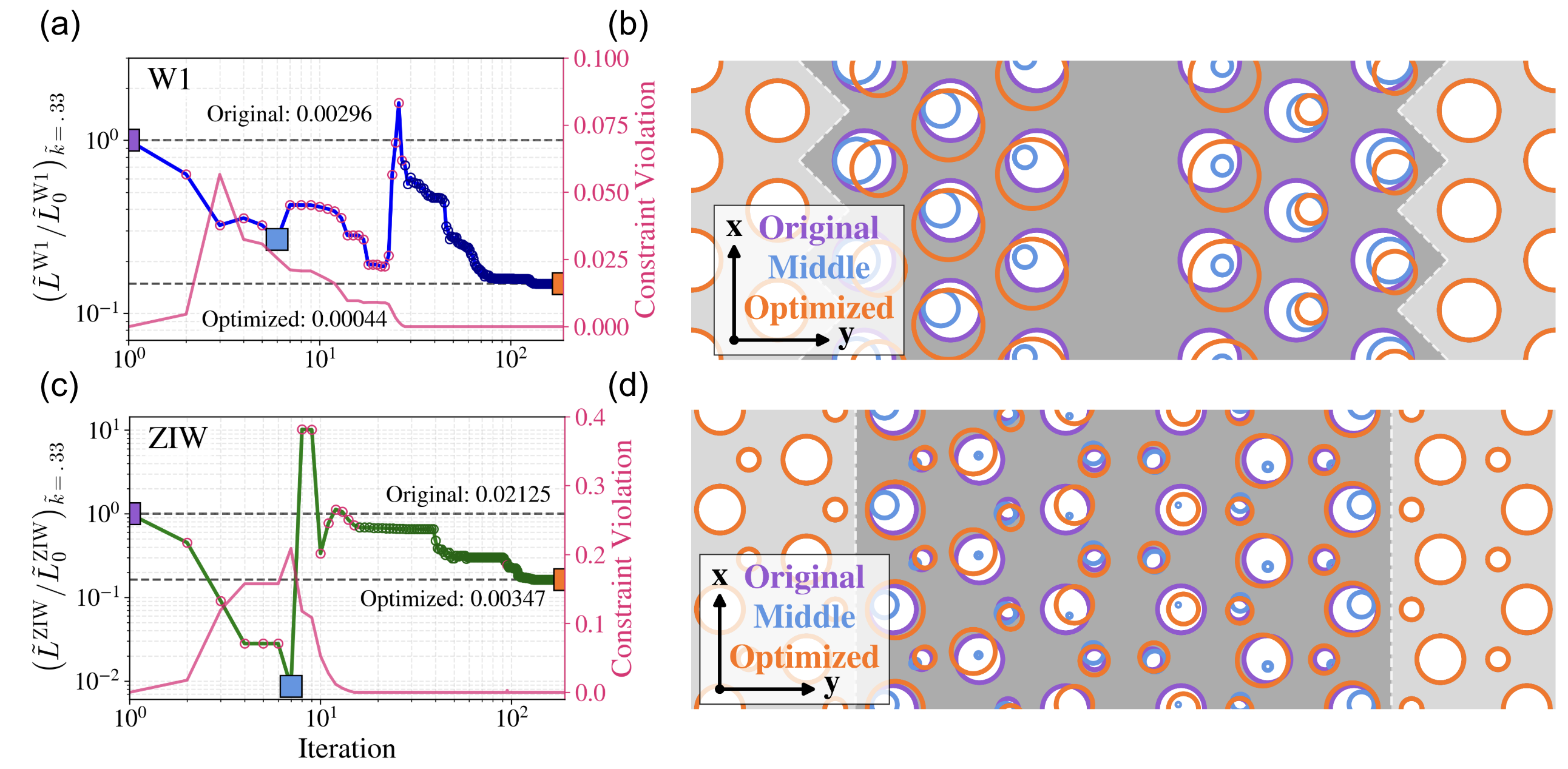}
    \caption{Backscatter loss 
    figures-of-merit versus iteration number in the inverse design process, for W1 and ZIW structures, normalized by the original design value. Panels (a) and (c) show the evolution of the loss figure-of-merit calculations during the inverse design processes, for the W1 and ZIW waveguides, respectively. In addition, the pink curve shows the maximum value that a constraint is violated by. This is primarily at the beginning of the optimization process, where the constraints are not strictly enforced. Panels (b) and (d) show the hole positions of the original, middle and optimized designs in purple, blue, and orange respectively. The light propagates along the $x$ direction in the center of the plot. The darker gray regions indicate the holes that were allowed to move during optimization. 
    }
    \label{fig:decrease}
\end{figure*}

\subsection{Inverse design process} \label{sec:inverse}
The premise of inverse design is to treat certain design parameters, $\bm x$, as inputs to a function that gives the figure of merit for the device, and then to minimize the function under designated constraints. 
Fundamentally, this is an optimization problem written as:
\begin{align}
    &\min_{\bm x}&\tilde L_{\tilde k}^{\text{W1}/\text{ZIW}}(\bm x)&, \nonumber \\&\text{subject to}&g_i(\bm x)\le0,
\end{align}
where the functions $g_i$ represent constraints placed on the final design, $\tilde L_{\tilde k}^{\text{W1}/\text{ZIW}}$ is the function giving the figure of merit, and $\bm x$ are the design parameters. Equality constraints can also be used, but are not used in this paper. There are a variety of ways to find good solutions within the design space, but we have chosen to use gradient based optimization as it is known to be effective~\cite{Deng2024,Nussbaum2021,PhysRevA.106.033514}. Gradient based inverse design is only possible since the derivatives can be efficiently calculated for both GME and the formula for backscattering, Eq. \eqref{eq:backFull}~\cite{Minkov2020}. 

A typical gradient based optimization routine can be used to minimize the figure of merit of interest. This allows for a much more efficient search of the design space than conventional brute force methods. We define the figure of merit as,
\begin{equation}\label{eq:fom}
    \tilde L_{\tilde k}^{\text{W1}/\text{ZIW}}(\bm x)=\frac{\langle\alpha_{\text{back}}\rangle_{\tilde k}}{n_g^2(\tilde k)},
\end{equation}
with $\langle\alpha_{\text{back}}\rangle$ defined by Eq.~\eqref{eq:backFull}.We do this because, in principle, losses can be reduced in two ways: (1) by reducing the field strength at the hole edges, where any imperfections are found; and (2) by reducing the group index, which decreases not only the losses (c.f. Eq.~\eqref{eq:backFull}) but also the strength of light-matter interactions within the waveguide \cite{MangaRao2007}. Optimizing $\langle \alpha_{\text{back}}\rangle$ directly results in a large reduction in $n_g$. Therefore, we remove the group index-dependence from the figure of merit to ensure that loss is primarily reduced due to the first mechanism. The gradient information can be obtained through the use of the Python automatic-differentiation package \textit{Autograd}. 

When optimizing the geometry of the PCW structure, we would like to preserve certain properties of the design to ensure that the results remain usable. For instance, the 
waveguide mode of interest should stay within the photonic band gap, be single-mode, and the structure must remain feasible to manufacture. This imposes constraints on the optimization problem, some of which are nonconvex and highly nonlinear. This is handled through the \textit{trust-constr} method from {\it Scipy}, which has been shown to be effective for complex constraints and large numbers of design parameters \cite{Zhang2021}. Below, we provide a brief and simplified description of how the method handles such constraints.

The \textit{trust-constr} method handles constraints in two different ways. At the start of each iteration, it defines a barrier subproblem 
\begin{equation}
    \phi(\bm x) = f(\bm x_t)-\mu_t\sum_i\ln(-g_i(\bm x_t)),
\end{equation}
which it attempts to solve such that the constraints are violated by no more than a barrier tolerance. Here, $f(\bm x_t)$ and $g_i(\bm x_t)$ are the figure-of-merit and constraints for the $t^{\text{th}}$ iteration, respectively, and $\mu_t$ defines the contribution of the barrier term. Once a solution within the barrier tolerance has been found, the step is either accepted or rejected based on the merit function, defined as
\begin{equation}
    \Psi(\bm x_t)= f(\bm x_t)+\rho\sum_i\max(0,g_i(\bm x_t))^2,
\end{equation}
where $\rho$ controls the relative strength of the two terms. 

If the merit function has decreased significantly from the previous iteration, then the step is accepted; otherwise, it is rejected. Then the process is repeated with the barrier tolerance and $\mu$ being reduced. Importantly, due to the barrier tolerance and the nature of the merit function, it is common to find intermediate steps that violate constraints, but, as long as the constraints are satisfied by the end of the optimization, the optimization can still be successful. Examples of this can be seen in Fig.~\ref{fig:decrease}, panels (a) and (c). Throughout this paper, we use a barrier tolerance of $0.1$, $\mu=0.1$, and $\rho=1$.

\section{Numerical Results and Improved Designs for Lower Loss} \label{sec:results}

\begin{figure}
  \centering
  \includegraphics[width=\columnwidth]{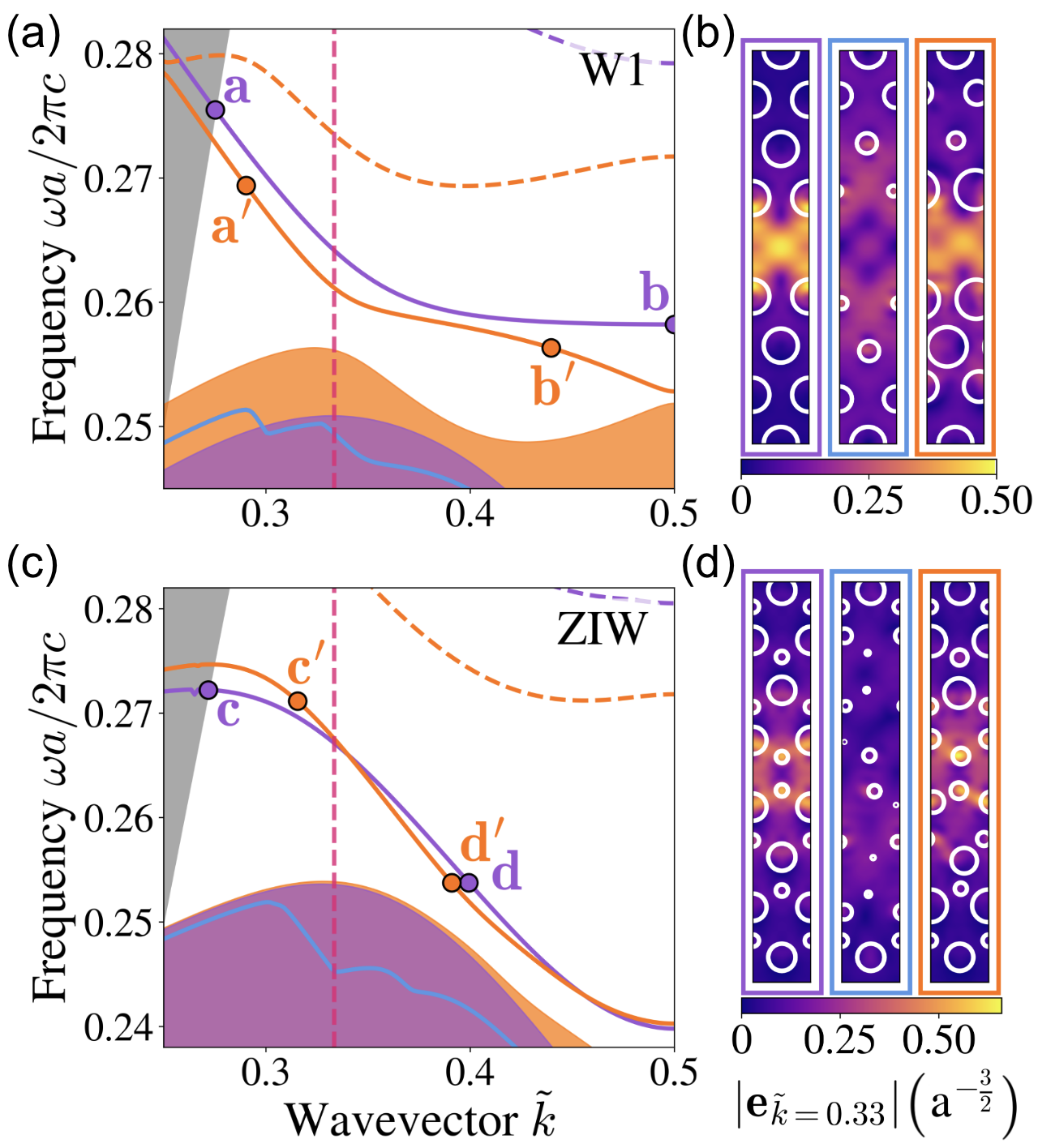}  
  \caption{
  Band structure and selected modes for 
  original and improved designs, as well as an example of one falling outside our constraints.
  Panel (a) shows the overlaid band structures of the W1 waveguide before (purple), in the middle of (blue), and after (orange) optimization. The gray region is the light line, the darker regions below are the respective bulk modes (the midpoint not shown), solid lines are the bands of interest, and dashed horizontal lines are additional, higher-order bands within the band gap. The pink dashed vertical line is the $\tilde k=0.33$ point that optimizations were preformed. The points $a$, $b$, $a'$ and $b'$ mark the edges of the single moded region for the original and final design, respectively. Panel (b) shows the electric field in the middle of the slab before, in the middle of, and after optimization (left to right). Panels (c) and (d) are corresponding plots, but for the ZIW waveguide designs.}
  \label{fig:results}
\end{figure}

\begin{figure}
  \centering
  \includegraphics[width=.95\columnwidth]{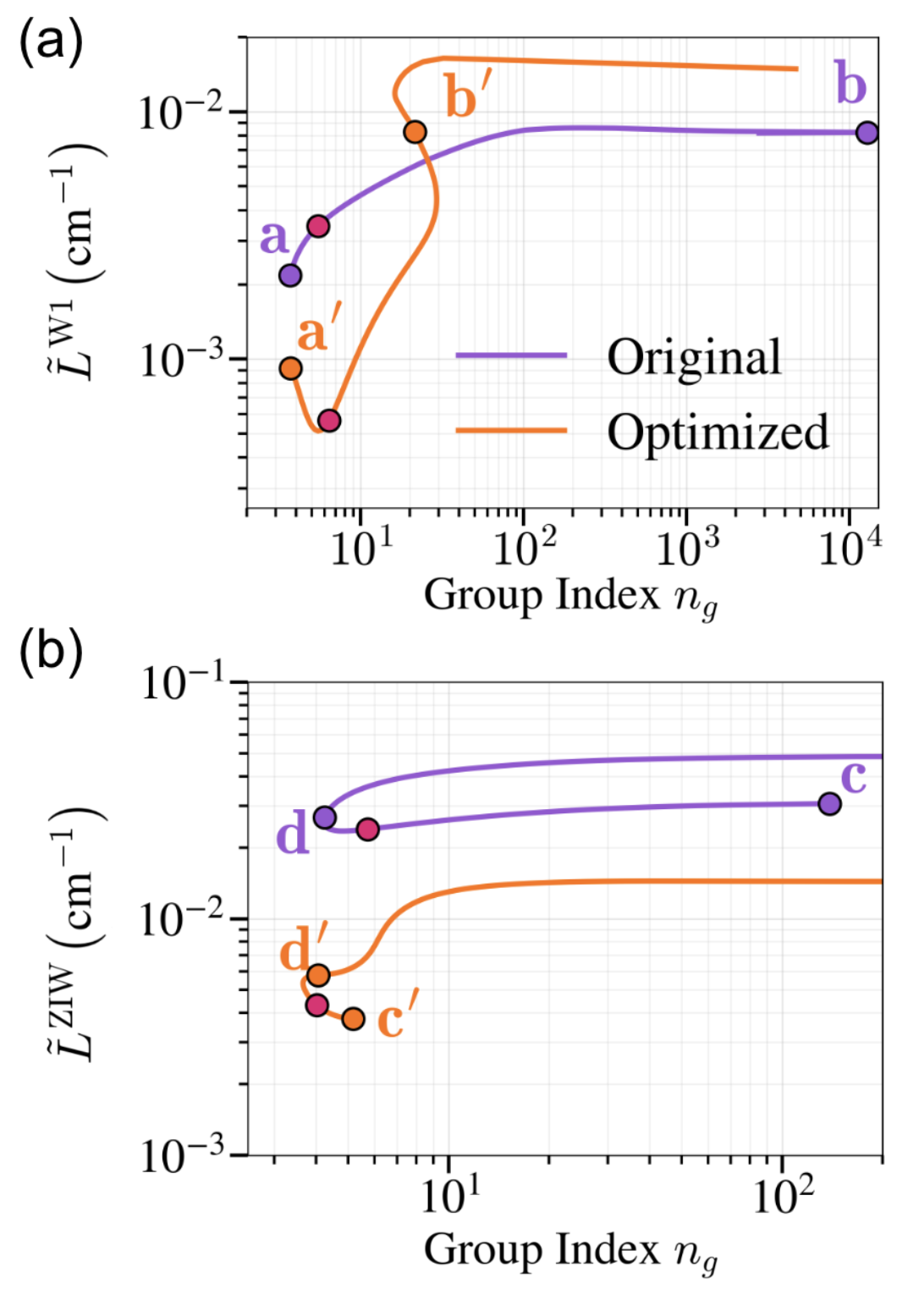}  
  \caption{Backscatter loss 
  figures-of-merit (loss with the group index component removed) versus group index values, for the original and optimized W1 (a) and ZIW (b) waveguides. The annotated points correspond to the edges of the single moded region shown in Fig.~\ref{fig:results}, and the pink points correspond to the $k$-point being optimized.}
  \label{fig:Losses}
\end{figure}

For our first calculations, we ran a set of numerical optimizations on the W1 and ZIW waveguides, as shown in Fig.~\ref{fig:diagram}, using Eq.~\eqref{eq:fom} as the figure-of-merit, computed at a fixed $\tilde k'=k'_x(2\pi/a)=0.33$. In our case, we wish to minimize this figure of merit.

In the computations, we allow the $x$-$y$ positions and radii of the holes in the first three unit cells on each side of the waveguide to change independently, i.e., the dark-gray region shown in Figs.~\ref{fig:decrease}(b,d). This limited number of holes was chosen because it already provides a significant impact on backscattering while keeping GME numerically stable and maintaining a well-defined photonic bandgap \cite{Mann:13, pattersonDisorderinducedIncoherentScattering2009}. 

We also highlight that
the optimization problem was sensibly constrained by ensuring the structure 
is both usable and compatible with fabrication tolerances; specifically, we use a minimum radius of 27.5 nm and a minimum distance between holes of 40 nm \cite{PhysRevA.106.033514}.Additionally, to ensure that the mode being optimized remains single-mode, we impose three different conditions on the mode of interest, $\omega_{\tilde k',i}$, where $i$ indexes the band in terms of increasing frequency. The first two conditions are $\omega_{\tilde k',i}+b<\omega_{\tilde k',i+1}$ and $\omega_{\tilde k',i}-b>\omega_{\tilde k',i-1}$, for all $k$ and $b=0.001 \ (\omega a/2\pi c)$ is a bandwidth. This forces the mode of interest to remain separate in frequency with higher-order modes within the band gap and the continuum of modes at the band edge. In addition, we impose the condition $\omega_{\tilde k_1,i}>\omega_{\tilde k_2,i}$ for all $\tilde k_1,\tilde k_2<0.5$ with $\tilde k_1<\tilde k_2$. This ensures that the $i^{\text{th}}$ band remains single-moded. In order to practically implement these conditions, we define constraints using a limited number of $k$-points. We found that using five separate $k$-points at $\tilde k=0.28,0.3125,0.33,0.416$ and $0.5$ resulted in stable results that obeyed the stated conditions. Without these constraints, the final design could have overlapped modes in frequency, leading to additional losses.



\begin{figure}
  \centering
  \includegraphics[width=\columnwidth]{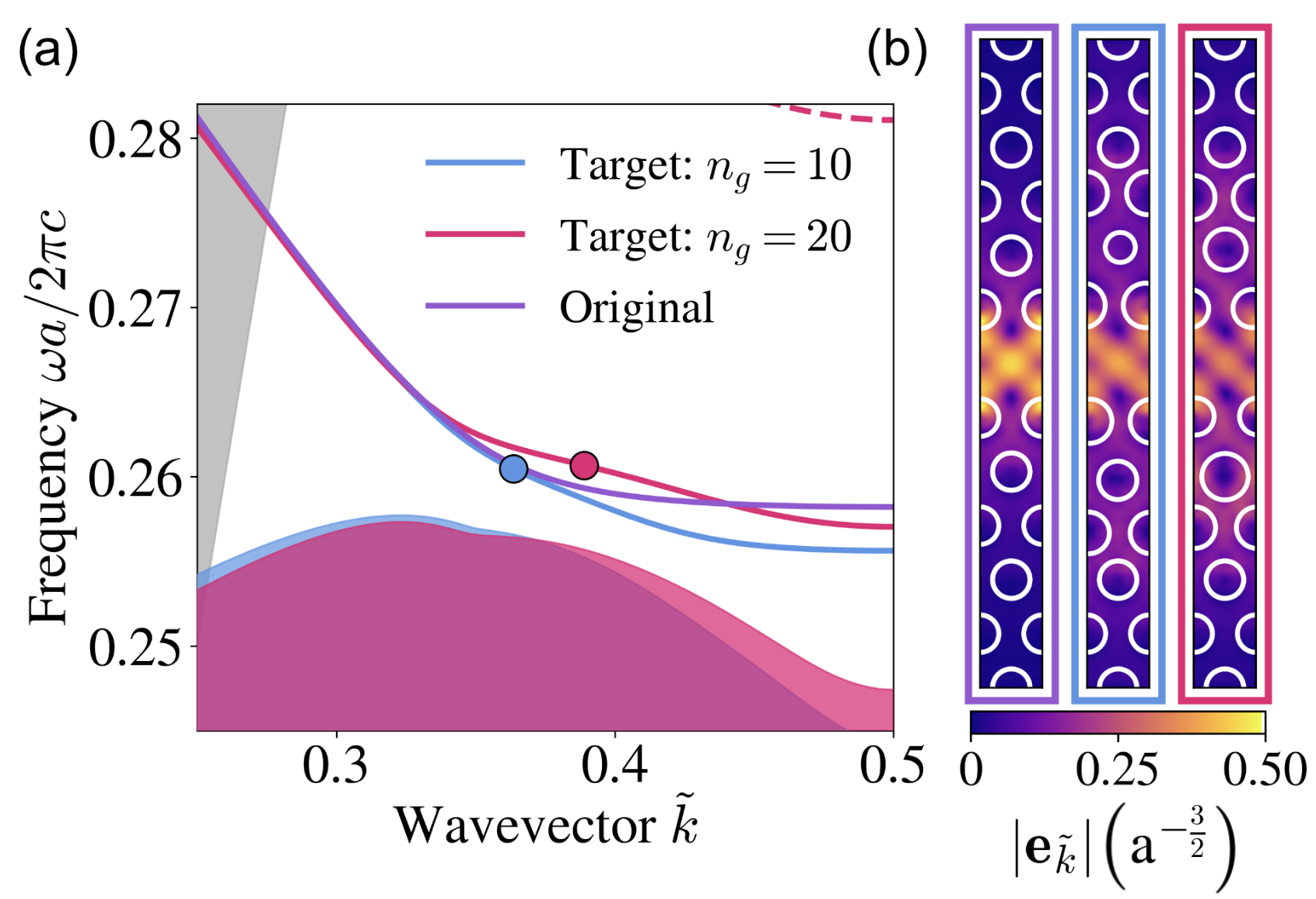}  
  \caption{Example inverse designs 
  for a contained group index
  of ten and twenty, for the W1 PCW.
  Panel (a) shows the band structure for a W1 PCW {\it before} (purple) and {\it after} an optimization targeting a group index of ten (blue) and twenty (magenta) at the $\tilde k=0.36$ and $\tilde k=0.39$ points, respectfully. The solid lines are the bands of interest, the gray region is the area above the light line, the solid region is the bulk modes outside the band gap, and the dashed lines are higher order modes within the band gap. The markers highlight the previously mentioned 
  $\tilde k$ points. Panel (b) shows the Bloch-mode electric field and hole locations for the original and optimized structures at the $\tilde k=0.36$ and $\tilde k$ points previously stated, respectively.}
  \label{fig:ngResults}
\end{figure}

The optimization path and the initial (purple), middle (blue), and final (orange) waveguide designs can be seen in Fig.~\ref{fig:decrease} and the corresponding band structures and fields are shown in Fig.~\ref{fig:results}. During optimization, the algorithm first allows the constraints to be slightly violated before gradually strengthening them, allowing for a more thorough exploration of the design space, as previously discussed. This results in the pink line show in 
Figs.~\ref{fig:decrease}(a,c). As an example of what causes this, the two order-of-magnitude reduction in loss seen in the ZIW waveguide is caused by the holes in the center being almost entirely removed, as can be seen in Fig.~\ref{fig:decrease}(d). This reduces backscattering, but the mode falls into the continuum of modes, resulting in a mode that is no longer localized to the center of the waveguide, as shown in Figs.~\ref{fig:results}(c) (W1) and (d) (ZIW). Additionally, many overlapping modes appear within the band gap and have been removed for clarity. The W1 waveguide design violates constraints in a similar way, as can be seen in Figs.~\ref{fig:decrease}(a,b) and Figs.~\ref{fig:results}(a,b). As the optimization progresses, the constraints are strengthened, resulting in a significant bandgap and mode localized at the center of the waveguide for both designs.

For the W1 case, 
Fig.~\ref{fig:Losses}(a)
summarizes the loss results. At the $\tilde k=0.33$ point, which we optimized on, the backscattering losses improve from $\langle\alpha_\text{back}\rangle_0^{\text{W1}}=0.13$ cm$^{-1}$ to $\langle\alpha_\text{back}\rangle_f^{\text{W1}}=0.027$ cm$^{-1}$ and $\tilde L_0^{\text{W1}}=0.003$ cm$^{-1}$ to $\tilde L_f^{\text{W1}}=0.00044$ cm$^{-1}$.
The group index value changes from $5.44$ to $6.26$ at the $\tilde k=0.33$ point through optimization. Additionally, the losses decrease across a broad range of $n_g$ values, while The bandwidth reduces from $21$ THz to $14.7$ THz. One feature lost after optimization is the absence of large $n_g$ values within the band gap. This can limit the broad applicability of the design. We will explore an approach to circumvent this problem in the next section.


For the ZIW case,
Fig.~\ref{fig:Losses}(b)
summarizes the loss results. 
At the $\tilde k =0.33$ point, which we optimized for, the backscattering losses improve from $\langle\alpha_\text{back}\rangle_0^{\text{ZIW}}=0.834$ cm$^{-1}$ to $\langle\alpha_\text{back}\rangle_f^{\text{ZIW}}=0.07$ cm$^{-1}$, and $\tilde L_0^{\text{ZIW}}=0.022$ cm$^{-1}$ to $\tilde L_f^{\text{ZIW}}=0.0034$ cm$^{-1}$. The group index value, at $\tilde k=0.33$, changed from $5.68$ to $4$.
The bandwidth increases slightly, from $19.5$ THz to $21$ THz. Unlike the W1 waveguide design, the ZIW optimized design has a much smaller set of $n_g$ values within the bandgap, which can be desirable for designs that aim to limit dispersion across the band.

\begin{figure}
  \centering
  \includegraphics[width=\columnwidth]{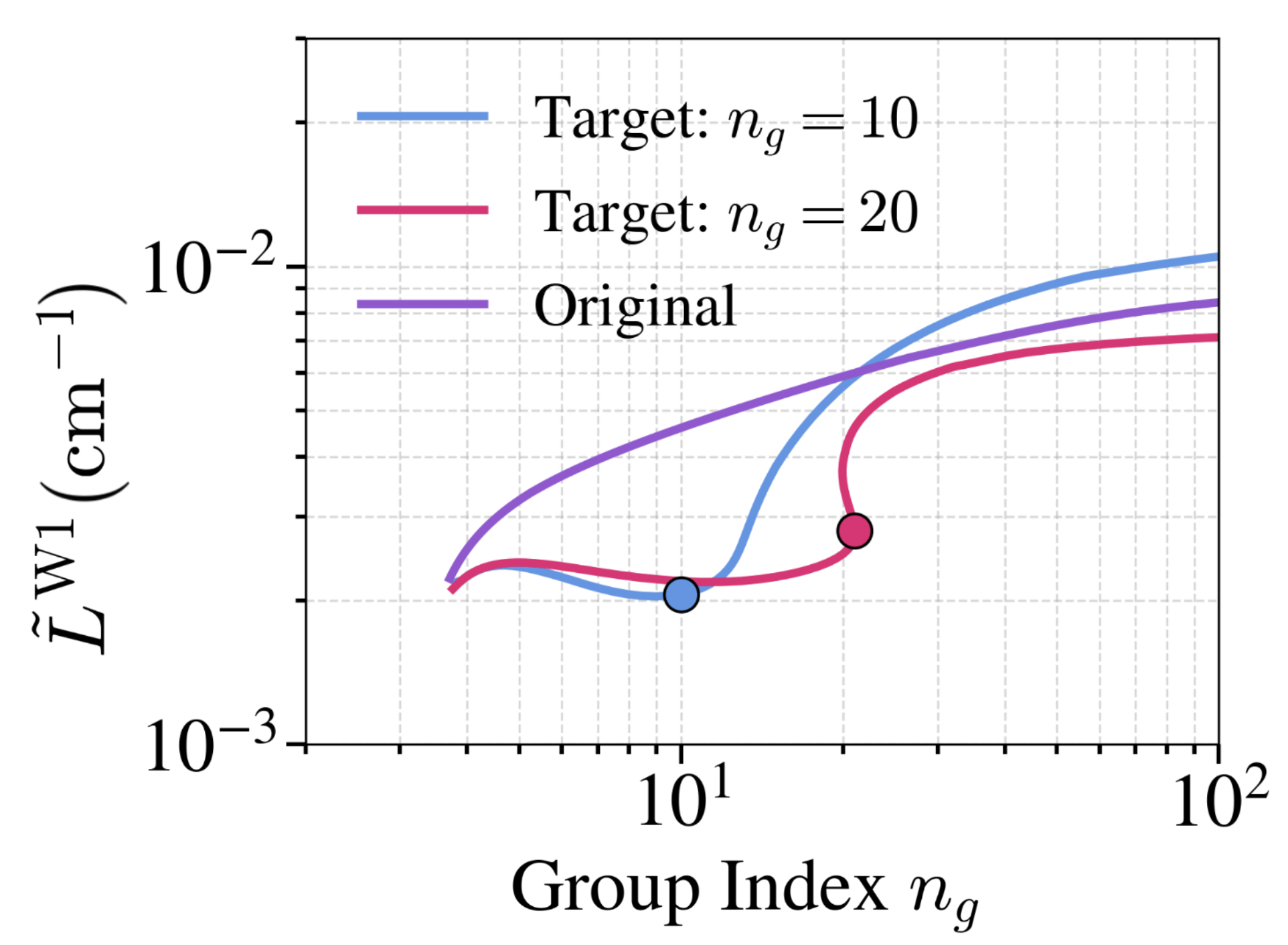}  
  \caption{Backscatter loss figure-of-merit 
  (recall this has the trivial group index removed), at each group index for the optimized W1 PCW, targeting low loss at the stated group index values. The markers label the group index location that the optimization was preformed on. The purple curve shows the original loss for the W1 PCW.
}
  \label{fig:ngLosses}
\end{figure}

\subsection*{Numerical Results for 
Group-Index-Constrained Loss Optimization}\label{sec:ngconstr}

If a particular application requires a lower loss mode at a larger (or specific) group index, this can be achieved by imposing an additional constraint on the group index. 
As an example, in Fig.~\ref{fig:ngResults} we show the resulting band structure for two optimizations of the W1 waveguide targeting a group index, $n_{g,\text{target}}$, of ten (blue) and twenty (magenta). To do this, we chose to optimize at the $\tilde k=0.36$ and $\tilde k=0.39$ points, respectively, since these points initially correspond to the desired group index value. We then implemented additional constraints on the group index itself, where $n_{g}>0.9\,n_{g,\text{target}}$ and $n_{g}<1.1\,n_{g,\text{target}}$. This limits the shift in the group index through optimization while not over-constraining, and thus limiting, the optimization. Besides these constraints and the choice of $\tilde k$ points, the problem is identical to the optimization in the previous section.

At the two $\tilde k$ points, the original W1 PCW design had losses of $\tilde L_{0,\tilde k=0.36}^\text{W1} = 0.00397$ and $\tilde L_{0,\tilde k=0.39}^\text{W1} = 0.00494$, where the final designs improved both by around a factor of two, to $\tilde L_{f,\tilde k=0.36}^\text{W1} = 0.00206$ and $\tilde L_{f,\tilde k=0.39}^\text{W1} = 0.0028$ respectively. Because of the constraints on the group index, both optimizations experienced little change in the group index. The broader relationship between the group index and the loss can be seen in Fig.~\ref{fig:ngLosses}.

The reason for the smaller reductions in losses through optimization for larger group indices can be seen in 
Fig.~\ref{fig:ngResults}(b), where the modes are distributed more evenly through the region surrounding the waveguide compared to the fields shown in Fig.~\ref{fig:results}(b). This is a characteristic of modes with larger group indices. Consequently, these modes are more challenging to optimize, since, regardless of the location of the hole, it will still overlap with a similar portion of the mode.

\subsection*{Effective Mode Volume and Enhanced Emission from Quantum  Emitters}\label{sec:applications}

For applications of PCWs in nonlinear quantum optics,
two key design metrics are effective mode volume 
(per unit cell) and Purcell enhancement of a single quantum emitter (or dipole). The effective mode volume is a measure of the distribution of the mode within the PCW and more generally quantifies the local field strength. It governs the strength of the light-matter interaction, with a smaller mode volume leading to stronger interactions \cite{MangaRao2007}. As the field maximum, assuming that the dipole's polarization is aligned with the mode, 
\begin{equation}
    V_{\text{eff}}=\frac{\int_{\text{cell}}d\bm r\epsilon(\bm r)|{\bf e}_k(\bm r)|^2}{{\rm max}[\epsilon(\bm r)|
    {\bf e}_k(\bm r)|^2]}.
\end{equation}
Since we are dealing with PCWs, this is the mode volume per unit cell, and it can be used to connect to the usual Purcell formula \cite{Hughes2004,MangaRao2007,RevModPhys.87.347}.
Although this mode volume definition is often used, the effective mode volume more generally depends on both the emitter’s polarization and position \cite{PhysRevA.106.033514}, which we use below.


The Purcell factor is a measure 
of the enhanced spontaneous emission rate of a dipole in a structured medium
relative to a homogeneous medium~\cite{MangaRao2007}. For embedded 
quantum dipole emitters (such as quantum dots), which are treated as point dipoles, it is defined as 
\begin{equation}
    \text{PF}({\bm r}_0)=\frac{3\pi c^2 an_g|{\bf e}_k({\bm r}_0)\cdot \hat{\bm n}|^2}{\omega_d^2\sqrt{\epsilon_b}},
\end{equation}
where the quantum emitter, located
at position ${\bm r}_0$,
is assumed to be in the slab medium (with dielectric constant $\epsilon_b$),
has a dipole along the $\hat{\bm n}$ direction, and an emission frequency $\omega_d$. When the quantum emitter's dipole is aligned with the electric field, the Purcell factor can be written as
\begin{equation}
    \text{PF}({\bm r}_{0})=\frac{3\pi c^2an_g}{V_{\text{eff}} \omega_d^2\epsilon_b^{3/2}},
\end{equation}
making it clear that the enhancement depends on both the group index and the effective mode volume. 

For our optimized designs to be useful in applications involving light matter interactions, we must ensure that the mode volume and Purcell enhancements remain unaffected by the optimization process. Looking at how the minimum mode volume changes for the designs shown in Fig~\ref{fig:results}, we find that for the W1 waveguide, the mode volume increases from $0.28\  [(\lambda/n)^3]$ to $0.34\  [(\lambda/n)^3]$, where $\lambda$ 
is the wavelength at the $\tilde k=0.33$ point and $n=3.46$ is the index of refraction of the slab, or  $0.007$ [$\mu \text{m}^3]$ to $0.0086\ [\mu \text{m}^3]$. While for the ZIW waveguide, the mode volume decreased from 
$0.23\  (\lambda/n)^3$ to $0.15\  (\lambda/n)^3$, or $0.0054$ $(\mu \text{m}^3)$ to $0.0036\ (\mu \text{m}^3)$. Additionally, there is a large electric field within one of the air holes of the ZIW waveguide design as can be seen in Fig.~\ref{fig:results}(d), which, for example, can be used for applications in all optical trapping. The maximum Purcell enhancement within the W1 waveguide is originally $1.32$ and changes to $1.17$ while the ZIW waveguide starts with a Purcell enhancement of $1.60$ which decreases to $1.67$. 

Although we find that the mode volume and Purcell enhancement for the W1 waveguide design is slightly worsened through the optimization process, we find a relatively equivalent improvement in the mode volume and Purcell enhancement for the ZIW waveguide. Since we find alternate effects from the two waveguide designs, it appears that decreasing backscattering loss does not have an inherent impact on the mode volume or Purcell enhancement. This may seem counterintuitive at first since, in order to decrease loss, we are reducing the strength of the mode, but the key is where the mode amplitude is being decreased. The loss depends only on the strength of the field at the edge of the hole, while the mode volume and Purcell factor depend on the strength of the electric field at the position of the interaction. Finally, the mode structure of each of the final designs is no longer symmetric, so to couple into the waveguide, a specialized coupler may be needed.


\section{Conclusions} \label{sec:coclu}

\begin{figure}
  \centering
  \includegraphics[width=\columnwidth]{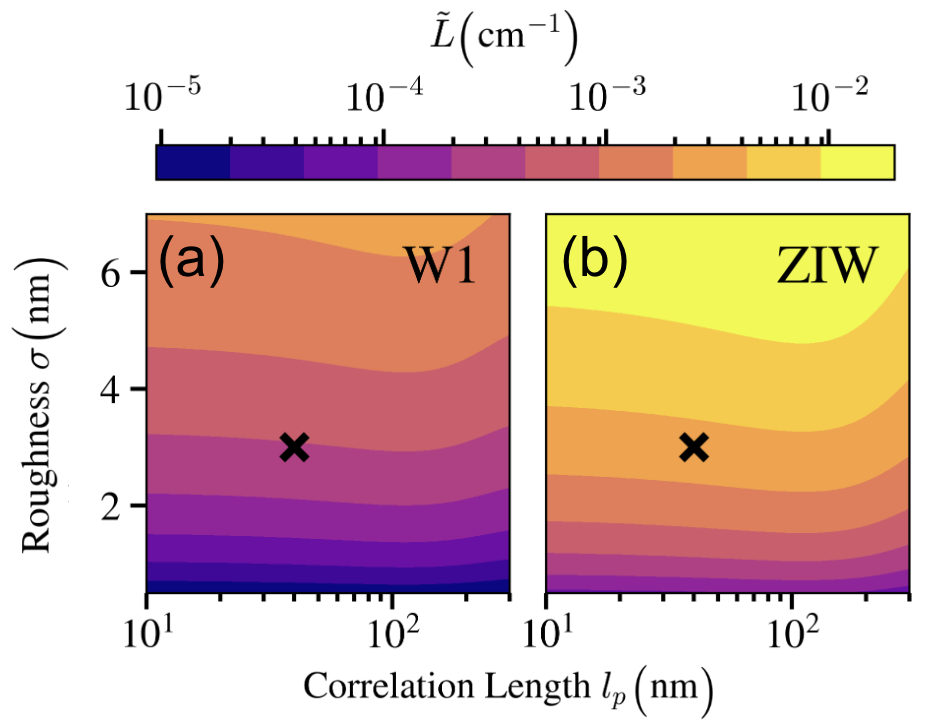}  
  \caption{Panels (a) and (b) show the effects 
  on the backscattering loss from changing the standard deviation ($y$-axis) and correlation length ($x$-axis) of the roughness around the hole edge, defined in Eq.~\eqref{eq:noise}, for the optimized W1 and ZIW waveguides in Fig.~\ref{fig:results}, respectively. The black crosses are located at the values we used above.}
  \label{fig:noise}
\end{figure}

We have shown how disorder-induced backscattering losses can be significantly reduced using inverse design by leveraging the speed of: 
(i) a physics-based analytical mode formula,
(ii) the GME method to obtain 
3D modes and dispersion,
(iii)
automatic differentiation, and 
(iv) effective use of existing constrained optimization algorithms.

The reduction in losses could have 
a significant impact on the performance of practical PCWs across various applications in photonics.  The W1 and ZIW waveguides showed a reduction in loss of 6.8 and 6.1 times respectively, compared to previous methods which at most showed a decrease of a factor of two  \cite{Mann:13}. Special attention should be 
given to the improved W1-like designs since it is the lowest backscattering loss PCW, for its group index of $n_g=6.73$, to our knowledge, with $\tilde L_{\tilde k}^{\text{W1}}=0.00044$ cm$^{-1}$ and $\langle\alpha_{\text{back}}\rangle_{\tilde k}^{\text{W1}}=0.027$ cm$^{-1}$ ($-0.117$ dB/cm). Note that this is using the statistical parameters $\sigma=3$ nm and $l_p=40$ nm for the noise around the hole edge, which is what has previously been used in the literature (and compared with select experiments). 

The nature of fabrication noise of course depends on the exact fabrication set up used to produce the waveguide. If the noise parameters are not known for a specific fabrication, then they can either be measured using imaging or matched to the loss values measured. As can be seen from Fig.~\ref{fig:noise}, the backscattering value depends quadratically on $\sigma$, so reducing, or at very least knowing, the roughness is an important component of decreasing loss.  
In addition, this method can easily be extended to improve or constrain many other important figures of merit in the PCW. For example, the group index, mode volume, Purcell enhancements, and bandwidth can all be controlled or improved using this method, while
also keeping the losses low, as demonstrated with the group index constrained optimizations in Fig.~\ref{fig:ngResults} \cite{Nussbaum2021,PhysRevA.106.033514}. 

The inverse design problem here is challenging due to its extreme nonlinearity and nonconvexity, and while the method used here, {\it trust-constr} from {\it scipy}, has shown promising results, it is by no means the only method that could be used. Future work to investigate alternative gradient optimization methods and reinforcement learning methods in order to fully explore the space of possible designs is important. Our work is thus the first step in designing a system that can consistently and robustly design PCWs that can be targeted to specific applications.

\vspace{0.2cm}
\acknowledgements
 This work was supported by the Natural Sciences and Engineering Research Council of Canada (NSERC),
 the National Research Council of Canada (NRC),
 the Canadian Foundation for Innovation (CFI), and Queen's University, Canada.


\bibliography{refs}

\end{document}